\documentclass[
aps,
prd,
twocolumn,
showpacs,preprintnumbers,nofootinbib,
amsmath,amssymb,
aps,
superscriptaddress]{revtex4-1}

\usepackage{float}
\usepackage{graphicx}
\usepackage{dcolumn}
\usepackage{bm}
\usepackage[colorlinks=true]{hyperref}
\usepackage{color}

\definecolor{RED}{rgb}{1,0,0}

\usepackage{acro}

\begin{document}

\title{On the Potential Observation of False Deviations from General Relativity in Gravitational Wave Observations from Binary Black Holes}

\affiliation{Department of Physics, The Chinese University of Hong Kong, Shatin, N.T., Hong Kong}
\affiliation{Center for Relativistic Astrophysics and School of Physics, Georgia Institute of Technology, Atlanta, GA 30332}

\author{Peter T. H. Pang}
\email{thpang@phy.cuhk.edu.hk}
\affiliation{Department of Physics, The Chinese University of Hong Kong, Shatin, N.T., Hong Kong}
\author{Juan Calder\'on~Bustillo}
\email{juan.bustillo@physics.gatech.edu}
\affiliation{Center for Relativistic Astrophysics and School of Physics, Georgia Institute of Technology, Atlanta, GA 30332}
\author{Yifan Wang}
\affiliation{Department of Physics, The Chinese University of Hong Kong, Shatin, N.T., Hong Kong}
\author{Tjonnie G. F. Li}
\affiliation{Department of Physics, The Chinese University of Hong Kong, Shatin, N.T., Hong Kong}

\begin{abstract}
\noindent Detections of gravitational waves emitted by binary black holes allow for tests of General Relativity in the strong-field regime. In particular, deviations from General Relativity can be observed by comparing incoming signals to waveform templates that include parametrized deviations from General Relativity. However, it is essential that the General Relativity sector of these templates accounts for all predictable physics. Otherwise, missing physics might be mimicked by the ``beyond General Relativity'' sector of the templates, leading the analysis to report apparent deviations from General Relativity. Current parametrized tests implement templates that omit physical phenomena such as orbital eccentricity and higher-order modes. In this paper, we show how the omission of higher modes can lead to false deviations from General Relativity when these effects are strong enough. We study the extent of these deviations as a function of the mass ratio and the orbital orientation. We find that significant false deviations can arise when current tests are performed on signals emitted by asymmetric binaries whose orbital angular momentum is orthogonal to the line-of-sight. We estimate that the Advanced LIGO-Virgo network operating at its design sensitivity can observe false violations with a significance above $5 \sigma$ as often as once per year. Similar results are expected for other tests of General Relativity that use modified waveform models. Hence, we stress the necessity of accurate waveform models that include physical effects such as higher-order modes to trust future tests of General Relativity.

\end{abstract}

\maketitle

\section{Introduction}
\noindent Since its postulation in 1916 \cite{Einstein:1916vd}, General Relativity 
(GR) has successfully passed all consistency tests performed. Most of them have been testing gravity in its so called weak-field regime based on astrophysical observations \cite{bps,sse,cm}. It was not until the recent detection of gravitational waves (GWs) from binary black holes (BBH) \cite{Abbott:2016blz,Abbott:2016nmj,Abbott:2017vtc} and  a binary neutron star (BNS) \cite{TheLIGOScientific:2017qsa} that GR has started to be tested in its strong field regime. Since then, a variety of tests have been performed. As a result, bounds on the mass of the graviton improved by a factor of two previous existing ones \cite{TheLIGOScientific:2016src}. Tests of the dynamics predicted by GR have been performed by comparing the final black hole states predicted from the early and late inspiral emission. Moreover, the recent joint detection of GW170814 by the Advanced LIGO \cite{TheLIGOScientific:2014jea} and Virgo \cite{TheVirgo:2014hva} detectors, has allowed for the first direct measurement of the GW polarization, enabling to test the hypothesis that GWs are purely tensorial \cite{Abbott:2017oio,Isi:2017fbj}.
Finally, the detected signals have been compared to generalized, analytical inspiral-merger-ringdown waveform templates that include parametrized deformations with respect to GR. Although none of these tests have shown strong evidence for deviations from GR, $2\sigma$ deviations have been observed for some of these events. These deviations can be ascribed to the stochastic nature of the noise or any physics that has not been captured by our waveform models.

Most tests of GR using GW from BBH involve the comparison of the signal to waveform templates. As described above, in some cases these models explicitly include deviations from GR. In doing so, it is crucial that the GR piece (or ``sector'') of these templates, to which non-GR parameters (or a ``nonGR sector'') is added, captures all physics predicted by GR. Otherwise, it is possible that the omitted physics is mimicked by physics beyond GR, leading to apparent false deviations from GR. State-of-the-art tests of this kind use templates whose GR sector omits the impact of physical phenomena such as the orbital eccentricity of the binary and the higher modes of the GW emission.
In particular, current tests of GR implement underlying waveform models, or GR sector, that only consider the dominant $\ell = 2$ modes \cite{Hannam:2013oca}, omitting the impact of further higher-order modes (see Section~\ref{sec:gwfrombbh}). This is partly due to the fact that no waveform model accounting for the effect of both higher modes and generic spins is currently available. However, higher modes are known to have a large impact on the GW signal emitted by BBH with mass ratios $q=m_1/m_2 > 4$ and total mass $M=m_1+m_2 > 100M_\odot$ \cite{Pekowsky:2012sr,Bustillo:2015qty,Bustillo:2016gid,Varma:2014jxa,Varma:2016dnf,Capano:2013raa,Graff:2015bba}. 

In this work, we investigate how the impact of higher modes in the GW signal can be mimicked by physics beyond GR, leading to apparent deviations from GR if higher modes are omitted in the GR waveform model. Because no analytic waveform models including the effects of both spins and higher modes was available at the time this study was done, we restrict to the case of non-spinning binaries.
We note that, to date, no source in the region of the parameter space described above has been found \cite{Abbott:2017iws}. Moreover, it is known that matched-filter searches \cite{Usman:2015kfa}, which are in principle optimal once the morphology of the signal is known, have a poor sensitivity in this part of the parameter space. This is partly due to the fact that current templates used as filters omit the higher-order modes of the signals, and partly due to the strong background of non-Gaussian noise transients affecting the high mass region of the parameter space \cite{DalCanton:2017ala}. In fact, this region is currently better covered by morphology agnostic searches \cite{Klimenko:2015ypf,CalderonBustillo:2017skv}. However, the current development of enhanced matched-filter searches \cite{Harry:2017weg,Nitz:2017lco} and the expected increase in sensitivity of the Advanced LIGO detectors for future observation runs might enable the first observation of the sources considered in this study.

The rest of this paper is structured as follows. Section \ref{sec:gwfrombbh} provides a brief overview of higher-order modes and the situations in which these produce strong effects. Sections~\ref{sec:tgr} and \ref{sec:inference} introduce the parametrized test of GR and its implementation within the framework of Bayesian Inference. Section~\ref{sec:simulation} gives an overview of the simulated signals and recovery templates we use. In section \ref{sec:results} we show that omission of higher modes in the GR sector of our templates can lead to strong apparent deviations from GR. This is, in the case of highly asymmetric BBH whose orbital plane is not perpendicular to the line-of-sight. Finally, we provide an outlook for future gravitational-wave detections Sec.~\ref{sec:conclusions}.

\begin{figure}
\includegraphics[width=\columnwidth]{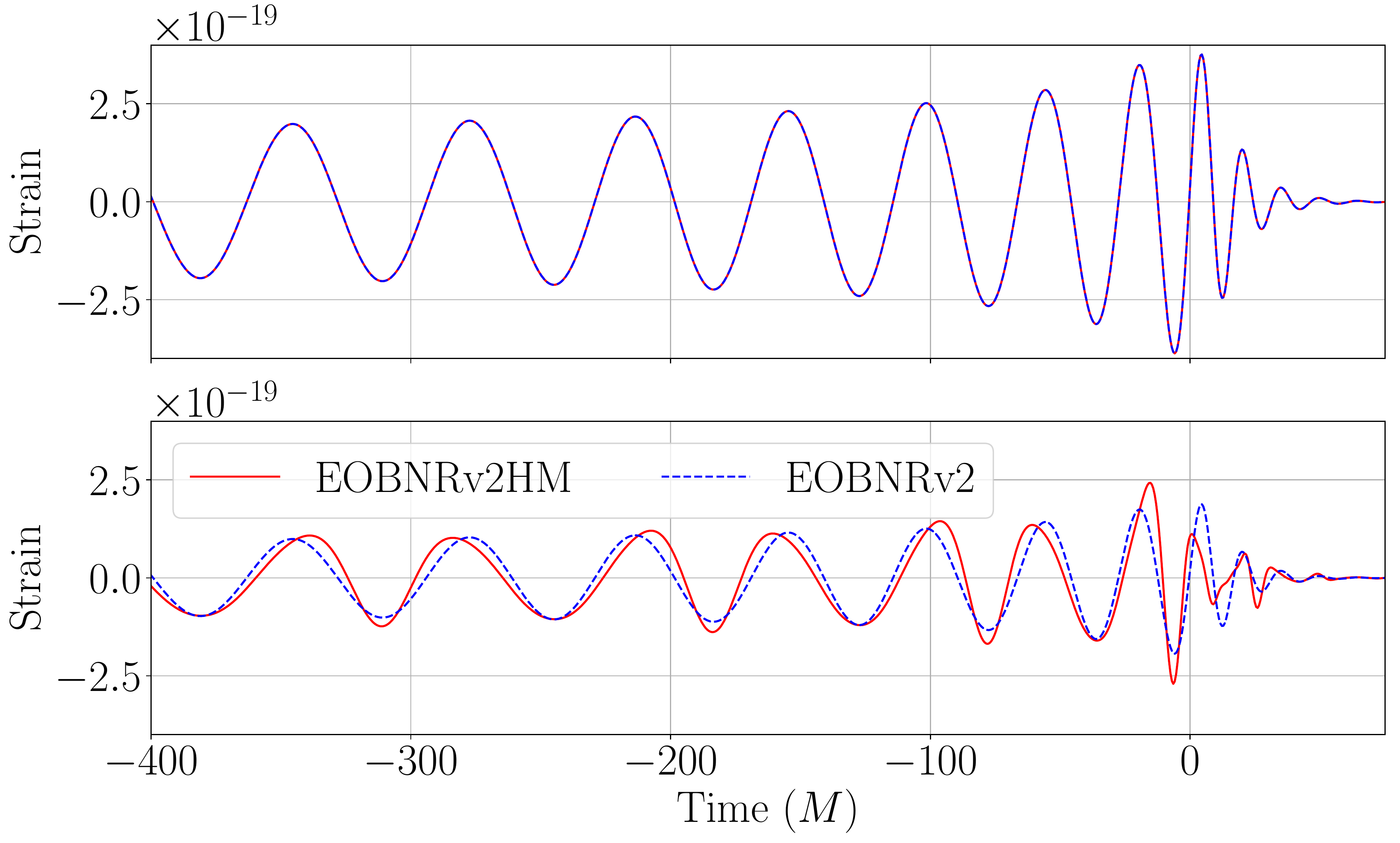} 
\caption[Effect of $q$ on higher modes]{\textbf{Impact of higher modes in a time domain waveform:}. The last cycles of a waveform from a non-spinning, $q=8$ binary (in geometrical units) with only the $(2,2)$ mode (EOBNRv2, blue dashed) and with higher-order modes included (EOBNRv2HM, red solid), when the source is face-on (top panel) and edge-on (bottom panel). There are no visible differences when the source is face-on, but higher modes introduce strong features near the merger-ringdown regime in the edge-on case.}
\label{ex:fig:waveforms}
\end{figure}

\section{Gravitational waves from binary black holes}\label{sec:gwfrombbh}
\noindent In GR, the emission from a non-eccentric BBH is characterized by a set of 15 parameters. The set of 8 parameters given by the masses $m_{1,2}$ and spins $\vec s_{1,2}$ of the individual black holes are known as the intrinsic parameters $\Xi$ of the binary.
Next, the sky location of the observer in the frame of the detector can be described in standard spherical coordinates $(r,\iota,\psi)$ with origin in the center of mass of the BBH. The axis $\iota=0$ is chosen to be orthogonal to the orbital plane, defined by $\iota=\pi/2$. In this framework, a BBH is said to be face-on or edge-on if the observer is located at $\iota=0$ or $\iota=\pi/2$, respectively. Finally, the time of coalescence $t_c$, the sky-location of the binary in the sky of the detector $(\theta,\varphi)$, and the polarisation angle $\psi_p$ of the signal complete the 15 parameters set.

The GW strain $h$ of a binary located at a luminosity distance $d_L$, as observed by a GW detector with antenna pattern $(F_+,F_\times)$ \cite{Jaranowski:1998qm} can be expressed as a combination of the two GW polarizations $(h_+,h_\times)$ as
\begin{align}
&h(\Xi,d_L,\iota,\psi,\theta,\phi,\psi_p;t-t_c)= \nonumber\\
&\quad \sum_{i=+,\times} F_i(\theta,\varphi,\psi_p) h_i (d_L,\iota,\varphi,\Xi;t-t_c).
\end{align}
Each GW polarization can be expressed as a superposition of gravitational wave modes $h_{\ell,m}$
\begin{equation}
\begin{aligned}
h_+ - ih_\times &= \frac{1}{d_L} \sum_{\ell\geq 2}\sum_{m=-\ell}^{m=\ell}Y^{-2}_{\ell,m}(\iota,\varphi)h_{\ell,m}(\Xi;t-t_c)\\
& \equiv h.
\end{aligned}
\label{gwmodes}
\end{equation}
For the case of non-precessing binaries, the above sum is always dominated by the $(\ell,m)=(2,\pm 2)$ modes during the inspiral part of the coalescence. The remaining modes, known as higher-order modes, have a sub-dominant effect and only contribute significantly to the GW signal during the last few cycles and merger of the binary. Notably, the amplitude of the higher-order modes grows with the mass ratio $q$, so that highly asymmetric binaries tend to have strong higher modes. Finally, the orientation of the binary also impacts the contribution of higher modes to the GW signal. The spherical harmonics $Y_{2,\pm 2}$ have their maximum at $\iota=(0,\pi)$ and minimum at $\theta=\pi/2$, while the rest of the harmonics behave in different ways, having their maxima close to $\iota=\pi/2$. This makes the $(2,2)$ mode especially dominant for face-on/off sources, while edge-on sources have a more important contribution from higher-order modes \cite{Berti:2007fi,Pekowsky:2012sr,Bustillo:2015qty}.

\section{Parametrized tests of General Relativity}
\label{sec:tgr}

\noindent In this study we will focus on a generic test of GR that focuses on measuring parameterized deviations deformation from the template waveform \cite{PhysRevD.85.082003}.
In particular, the method uses a closed-form expression for the waveform which is parameterized by a set of parameters $\left\{ p_i \right\}$.
For example, in the so-called \texttt{IMRPhenomPv2} model these parameters are given by inspiral PN parameters $\left\{ \varphi_0 \ldots \varphi_7 \right\}$ and $\left\{ \varphi_{5l},\varphi_{6l} \right\}$, phenomenological inspiral-merger $\left\{ \beta_0 \ldots \beta_3 \right\}$ and merger-ringdown $\left\{ \alpha_0 \ldots \alpha_5 \right\}$.
Deviations from GR, or the non-GR sector, are represented by relative deviations $p_i \rightarrow (1+\delta \hat{p}_i) p_i$.
In this so-called parametrized test, GW data is compared to waveforms $h$ with the goal of obtaining a probability distribution for $\delta \hat{p}_i$. Values of $\delta \hat{p}_i\neq 0$ indicate that the GR sector of $h$ can not recover the whole physics present in the incoming signal, therefore providing evidence that the signal contains physics beyond GR. In these sorts of analyses, it is essential that the GR sector of the sector ${h}_{\textrm{GR}}(t)$ includes all the physics predicted by GR. Otherwise, it is possible that GR physics missing in GR sector of the template could be absorbed by the non-GR one. This would lead to an apparent false violation of GR. 

Alternatively, the idea of the parameterized tests can also be re-expressed as
\begin{equation}
\begin{aligned}
{h}(t) &= {h}_{\textrm{GR}}^{\textrm{incomplete}}(t) + {h}_{\textrm{nonGR}}(t)\\
\end{aligned}
\end{equation}
where, in general
\begin{equation}
{h}^{\textrm{incomplete}}_{\textrm{GR}}(t) = {h}_{\textrm{GR}}(t) - {h}^{\textrm{missing}}_{\textrm{GR}}.
\end{equation}
Above, ${h}_{\textrm{GR}}(t)$ denotes an ideal GR sector accounting for all physics predicted by GR while ${h}^{\textrm{missing}}_{\textrm{GR}}(t)$ denotes the part of the full GR sector omitted by some incomplete GR sector ${h}^{\textrm{incomplete}}_{\textrm{GR}}(t)$. And ${h}_{\textrm{nonGR}}(t)$ denotes nonGR sector accounting for physics beyond GR.

Current parametrized tests of GR implement a GR sector that omits the higher-order modes of the GW emission \cite{Hannam:2013oca}. Denoting by $\Xi$ the set of intrinsic parameters of a given BBH, we can express this as:
\begin{equation}
\begin{aligned}
{h}_{\textrm{GR}}(t,\Xi) = & \sum_{(\ell \geq 2,|m|\leq \ell)}
Y^{-2}_{\ell,m}(\iota,\varphi)h_{\ell,m}(t,\Xi), \\
{h}^{\textrm{incomplete}}_{\textrm{GR}}(t,\Xi) = & \sum_{(\ell = 2,|m|\leq \ell)}
Y^{-2}_{\ell,m}(\iota,\varphi)h_{\ell,m}(t,\Xi), \\
{h}^{\textrm{missing}}_{\textrm{GR}}(t,\Xi) = & \sum_{(\ell \neq 2,|m| \leq \ell)}
Y^{-2}_{\ell,m}(\iota,\varphi)h_{\ell,m}(t,\Xi).
\label{eq:sectors}
\end{aligned}
\end{equation}
In the following, we will investigate how the higher modes, or missing physics ${h}^{\textrm{missing}}_{\textrm{GR}}(t,\Xi)$ omitted by the GR sectors ${h}^{\textrm{incomplete}}_{\textrm{GR}}(t,\Xi)$ of the templates $h$ can be mimicked by the physics beyond GR present in its nonGR sector ${h}_{\textrm{nonGR}}(t)$, leading to apparent violations of General Relativity. Our goal is to systematically study the strength of these violations as a function of the parameters $\vec{\lambda}=(\Xi;\iota,\varphi)$ of the binary. To this end, we will perform parametrized tests of GR on a large set of simulated signals with different physical parameters, which refer to as \textit{injections}, including higher modes. In order to prove that any violations we find are caused the presence of higher modes in the injections, and not by other possible differences between the injections and the templates, we repeat our analysis using a second injection set with identical physical parameters in which higher modes are omitted.

\section{Bayesian Inference}
\label{sec:inference}

\noindent The inner product of two real functions $a(t)$ and $b(t)$ is defined as 
\begin{equation}
(a|b)=4 \Re \int_{f_{\textrm{low}}}^{f_{\textrm{high}}} \frac{\tilde a(f)\tilde{b}^{*}(f)}{S_n(f)}df.
\end{equation}
Here, $\tilde a(f)$ is the Fourier transform of $a(t)$, $*$ denotes complex conjugation and $S_n(f)$ denotes the one sided power spectral density of the detector noise. Our analysis considers the predicted Advanced LIGO sensitivity curve, also known as the zero-detuned-high-power configuration \cite{AdvLIGOcurves}. We use lower and higher frequency cutoffs $f_{\textrm{low}}$ and $f_{\textrm{high}}$, of 20Hz and 4096Hz respectively.

The posterior $p(\vec\lambda|d)$ that a signal $h(\vec{\lambda})$ with parameters $\vec{\lambda}$ is embedded in a given signal $d$, is given by
\begin{equation}
p(\vec\lambda|d,\mathcal{H}) = \frac{p(d|\vec{\lambda},\mathcal{H})p(\vec{\lambda}|\mathcal{H})}{p(d|\mathcal{H})}.
\end{equation}
Assuming Gaussian noise, the likelihood $p(d|\vec\lambda)$ that a signal $d(t)$ with unknown parameters is the result of burying some signal template $h(\vec{\lambda})$ with parameters $\vec{\lambda}$ in such noise, is given by 
\begin{equation}
	p(d|\vec{\lambda}) \propto \exp{\left[-\frac{1}{2}(d-h(\vec{\lambda})|d-h(\vec{\lambda}))\right]}.
\end{equation}
In general, the vector of parameters $\vec\lambda$ consists on a combination of 15 physical parameters $\{\Xi;d_L,\iota,\varphi;\bar\iota,\bar\varphi,\psi_p;t-t_c\}$ and the parameters $\delta \hat{p}_i$ accounting for deviations from GR.
The shape and spread of the likelihood function along the parameter space follow from the signal present in the data and the power spectral density of the noise. In this study, our data $d$ will consist in simulated signals $h_{\textrm{inj}}$ injected in Gaussian noise $h$, generated according to some power spectral density $S_n$. In what follows, we will refer to our input data $h_{\textrm{inj}}$  as \textit{injections}, while we will refer to our recovery waveforms $h(\lambda)$ as \textit{templates}.

The parametrized test of GR can be further extended to a model selection problem. In particular, one needs to decide which of two models is a better description of our GW data: a model that only accounts for GR physics, or another one including physics beyond it. Model selection is a problem that is now well established in gravitational wave data analysis in the framework of Bayesian inference. Given a simulated signal $h_{\textrm{inj}}$ we want to compare two \textit{hypothesis}: ${\cal H}_0$ and ${\cal H}_1$. These assume respectively that data is well described by GR and that parametrized deviations from GR are needed for a good description of the data. The posterior probability of $\cal H$ being true is given by
\begin{equation}
p({\cal H}|d) = \frac{p(d|{\cal H})p({\cal H})}{p(d)},
\end{equation}
where the marginal likelihood $p({\cal H})$ is given by 
\begin{equation}
p(d|{\cal H}) = \int  p(\vec\lambda,{\cal H})p(d | \vec\lambda, {\cal H}) d\vec\lambda.
\end{equation}
The two hypothesis can be then compared by means of their odds that, given by 
\begin{equation}
{\cal O}^{1}_{0} = \frac{p(\mathcal{H}_{1}|d)}{p(\mathcal{H}_{0}|d)}=\frac{p(d|\mathcal{H}_{1})}{p(d|\mathcal{H}_{0})}\frac{p(\mathcal{H}_{1})}{p(\mathcal{H}_{0})} \equiv \mathcal{B}^{1}_{0}\frac{p(\mathcal{H}_{1})}{p(\mathcal{H}_{0})}.
\end{equation}
Above, $\cal O$ and $\mathcal{B}$ are the so-called \textit{odd ratio} and \textit{Bayes factor}, respectively. A result of ${\cal O}^{1}_{0} > 1$ indicates that ${\cal H}_1$ provides a better description of the data than ${\cal H}_0$. 

In this work, we want to show how inclusion of higher modes in the injections can lead to apparent deviations from GR when these are omitted in the templates. Since the effect of higher-order modes is only significant in the merger-ringdown stages of the binary \cite{Bustillo:2016gid,Bustillo:2015ova,Pekowsky:2012sr,Berti:2007fi}, we focus our analysis on these parts of the emission. Therefore, the set of extra GR parameters is reduced to those affecting these stages of the coalescence. These are commonly expressed as \cite{TheLIGOScientific:2016src}:
\begin{equation}
	\{\delta \hat{p}_n\} = \{ \delta\hat{\alpha}_2,\delta\hat{\alpha}_3,\delta\hat{\alpha}_4\}.
\end{equation}
The parametrised test will yield posterior distributions for each of the above parameters. We stress that values $\delta\alpha_i \neq 0$ indicate deviations from GR. In this situation, we can define the strength of a violation of GR as the distance between the median of the posterior distribution of a given nonGR parameter $\delta\alpha_i$ and $0$ measured in number of standard deviations. Finally, the odds ratio $O^{\textrm{nonGR}}_{\textrm{GR}}$ can be written as
\begin{align}
O^{\textrm{nonGR}}_{\textrm{GR}}&= \frac{1}{7}\frac{P(\mathcal{H}_{\textrm{nonGR}})}{P(\mathcal{H}_{\textrm{GR}})}(B^{\delta\hat{\alpha}_2\neq0}_{\textrm{GR}}+B^{\delta\hat{\alpha}_3\neq0}_{\textrm{GR}}+B^{\delta\hat{\alpha}_4\neq0}_{\textrm{GR}} \nonumber \\
&+B^{\delta\hat{\alpha}_2,\delta\hat{\alpha}_3\neq0}_{\textrm{GR}}+B^{\delta\hat{\alpha}_2,\delta\hat{\alpha}_4\neq0}_{\textrm{GR}}+B^{\delta\hat{\alpha}_3,\delta\hat{\alpha}_4\neq0}_{\textrm{GR}} \nonumber\\
&+B^{\delta\hat{\alpha}_2,\delta\hat{\alpha}_3,\delta\hat{\alpha}_4\neq0}_{\textrm{GR}})\\
&\equiv \mathcal{B}^{\textrm{nonGR}}_{\textrm{GR}}\frac{P(\mathcal{H}_{\textrm{nonGR}})}{P(\mathcal{H}_{\textrm{GR}})}.
\end{align}
We will say that evidence for deviations from GR have been found when either the posterior distributions for $\{\delta\hat{\alpha}_2,\delta\hat{\alpha}_3,\delta\hat{\alpha}_4\}$ exclude the 0 value or when the odds ratio is such that $O^{\textrm{nonGR}}_{\textrm{GR}} \gg 1$. In our study, we take the the prior preference between hypothesis to be $1$. Therefore $O^{\textrm{nonGR}}_{\textrm{GR}} = \mathcal{B}^{\textrm{nonGR}}_{\textrm{GR}}$. Finally, the sampling of the likelihood function over our 9 to 12-dimensional parameter space is carried using the Nested Sampling algorithm $\verb+lalinference_nest+$ \cite{2015PhRvD..91d2003V} implemented in LAL \cite{LALSimulation}.

\section{Waveforms}
\label{sec:simulation}

\noindent The true GW emission from a BBH contains, in general, higher modes. Hence, for our injections, we need a model that includes higher modes. By the time this study was performed, the only available inspiral-merger-ringdown waveform model including higher modes was the effective-one-body model for non-spinning BBH presented in \cite{Pan:2011gk} \footnote{This model is available in the LIGO LALSimulation library \cite{LALSimulation} and denoted as EOBNRv2HM}. We note however that models for aligned-spin binaries including higher modes are under active development \cite{London:2017bcn,COTESTA}. The model we use includes the $(\ell,m)=(2,\pm 1),(2,\pm 2),(3,\pm 3),(4,\pm 4)$ and $(5,\pm 5)$ modes of the GW signal and its reliability has been tested recently in the appendix of \cite{CalderonBustillo:2017skv}.

We want to demonstrate how the effect of higher modes in the GW waveform can be mimicked by physics beyond GR. Therefore, we need to demonstrate that any violations we find are indeed produced by the effect of higher modes. Hence, we repeat our analysis using a target waveform model that omits higher modes and check that no violations are observed. To this, we choose as injections the restriction of the previous model to the $(2,\pm 2)$ modes \footnote{This model is known as EOBNRv2 in the LALSimulation library.}. We note that there exists a large collection of models of the $(2,\pm 2)$ modes emitted by a BBH that we could have used, in principle \cite{Husa:2015iqa,Khan:2015jqa,Taracchini:2012ig,Taracchini:2013rva, Buonanno2013,Santamaria:2010yb}. However, we want to be sure that the only differences in the two analyses is strictly due to the presence of higher modes in one of the models, and not due to different modelings of the dominant $(2,2)$ mode.

Following recent studies of this kind \cite{TheLIGOScientific:2016src} our templates implement GR sectors computed by the phenomenological inspiral-merger-ringdown model for precessing coalescing binaries described in \cite{Hannam:2013oca,Khan:2015jqa}, to which an extra nonGR sector is added in the parametrized way described in section \ref{sec:tgr}\footnote{This model is known as IMRPhenomPv2 in the LIGO Algorithm Library and we will refer to it as PhenomP throughout the text}. We note that this model aims to account for generic spin effects. In particular, an approximate modeling of precession effects is implemented via a precession parameter $\chi_p$. Finally, and most relevant for our purposes, this waveform model contains all $\ell =2$  modes, omitting the higher-order modes. Hence, our templates implement an incomplete GR sector ${h}^{\textrm{incomplete}}_{\textrm{GR}}$ that can be expressed as in Eq.~\ref{eq:sectors}.

\begin{figure*}[hbt!]
\includegraphics[width=1\columnwidth]{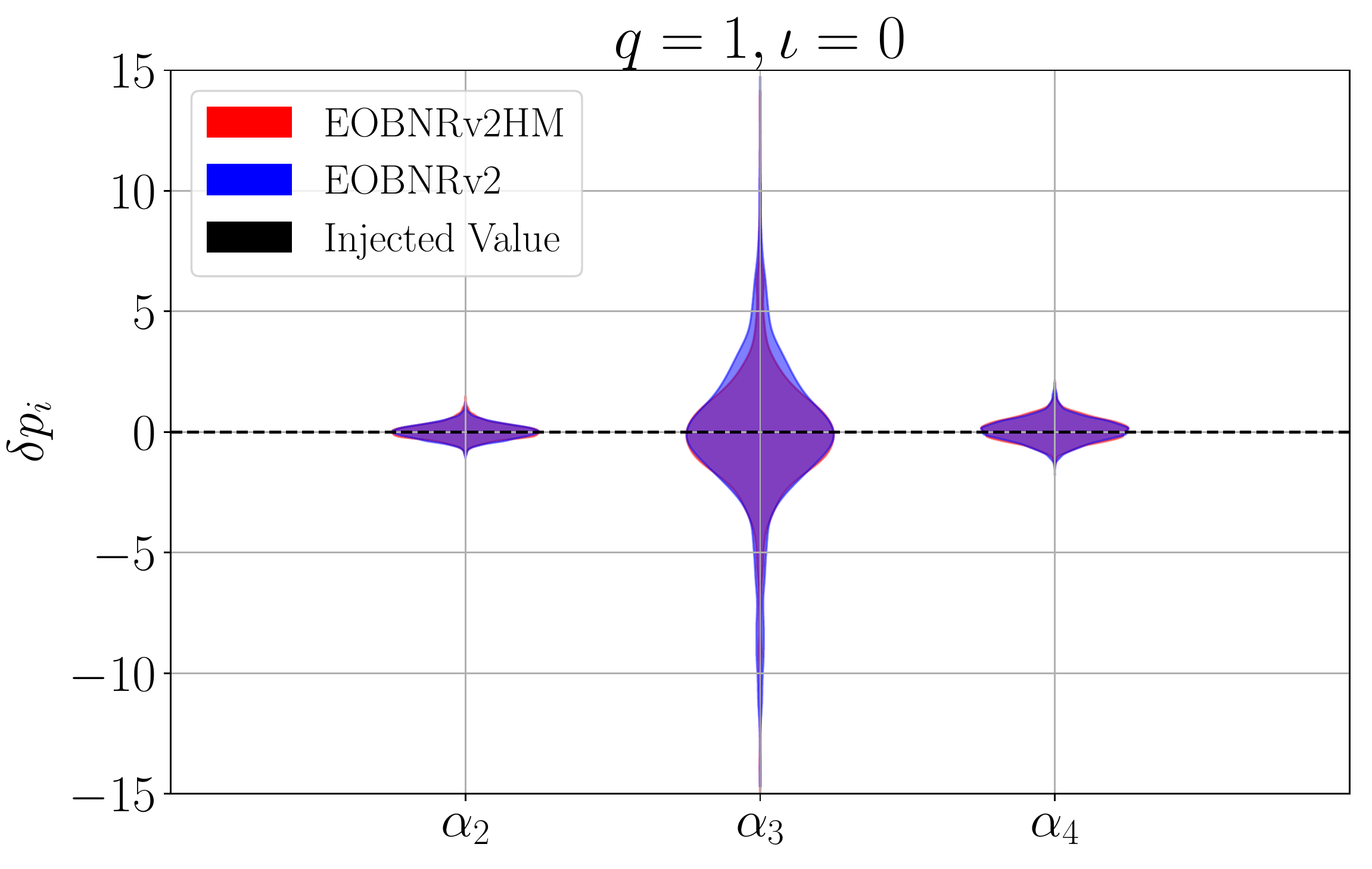}
\includegraphics[width=1\columnwidth]{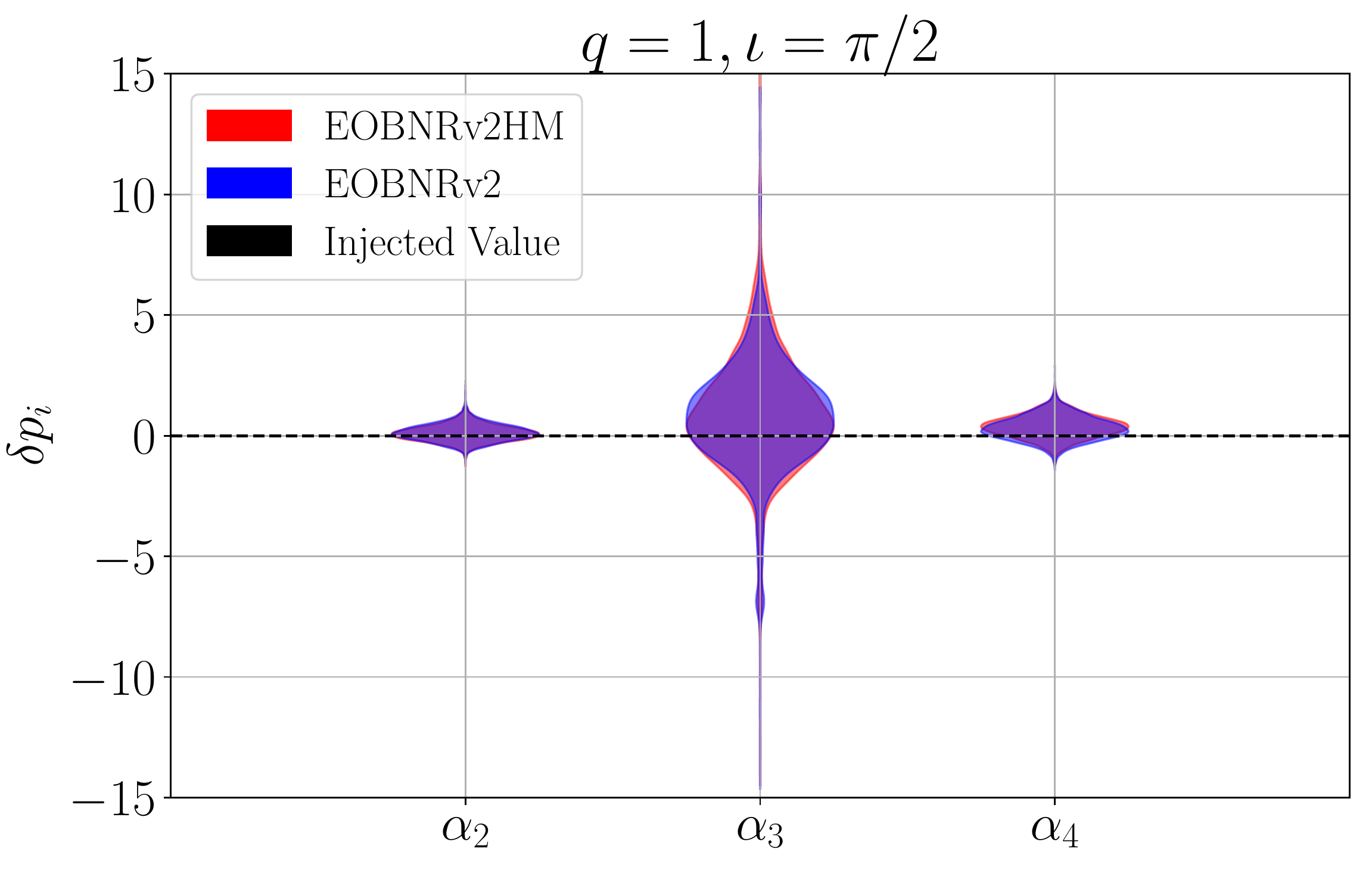}\\
\includegraphics[width=1\columnwidth]{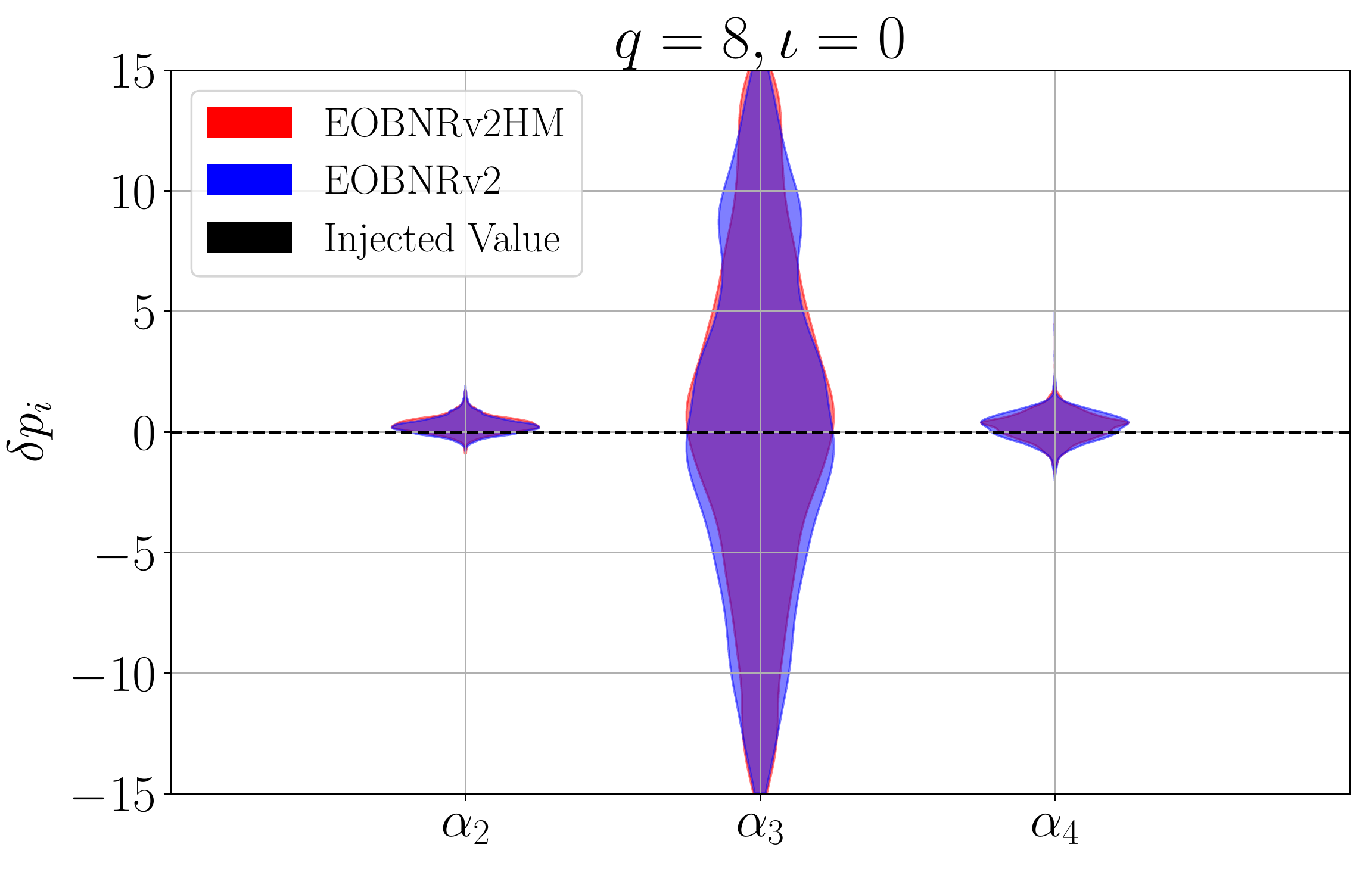}
\includegraphics[width=1\columnwidth]{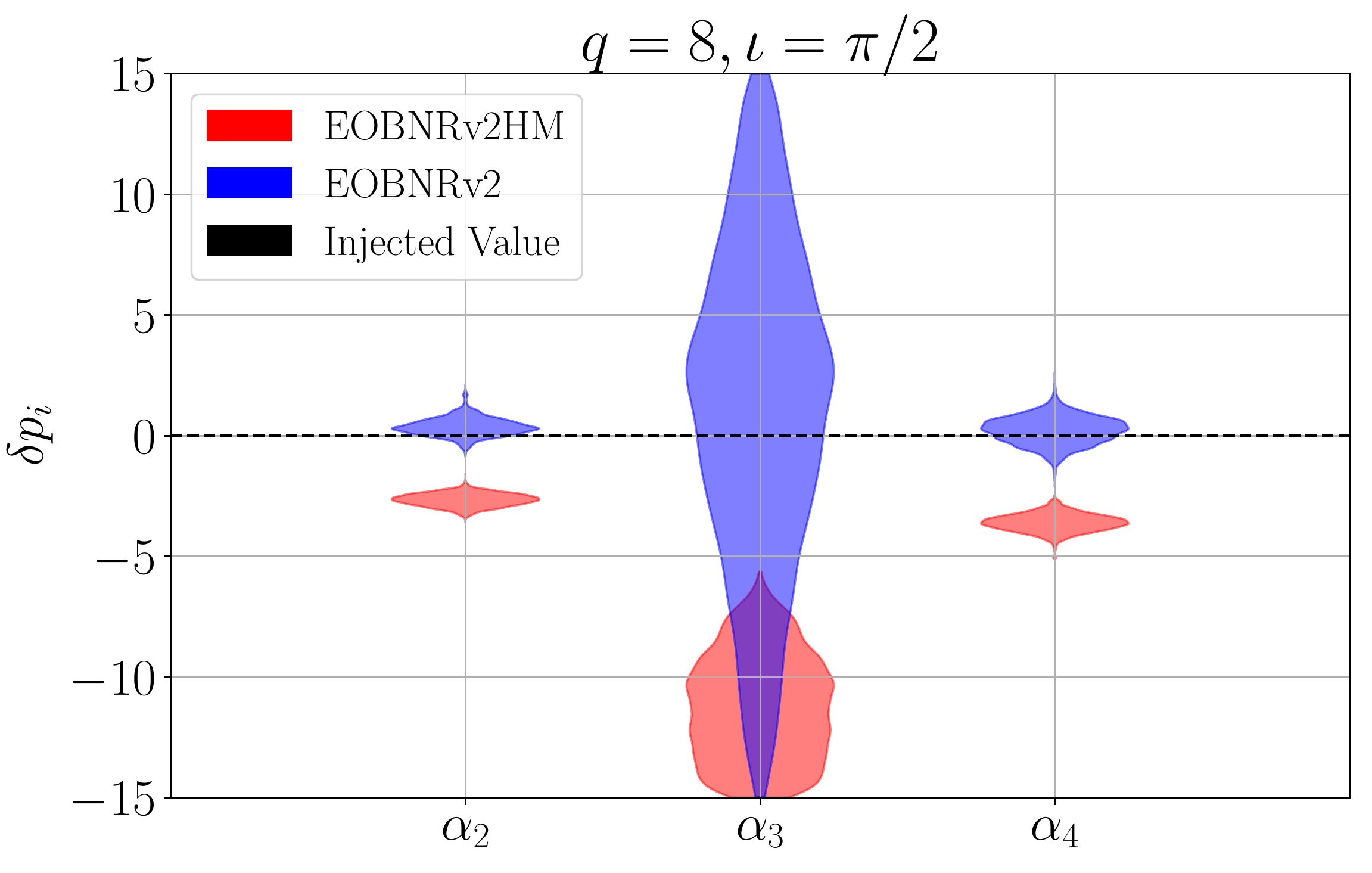}
\caption[Effect of q on TGR]{\textbf{Impact of higher modes on parameterized deviations from GR:} Posterior distributions of the nonGR parameters affecting the merger-ringdown emission ($\alpha_2$, $\alpha_3$ and $\alpha_4$) obtained for four different injections. These correspond to four BBH sources with total mass of $M=200M_\odot$, but different inclination angle $\iota$ and mass ratio $q$. In all cases, we inject signals including (red) and omitting (blue) higher modes. All injections have an optimal SNR $\rho_{\textrm{opt}} =15$. Note that significant deviations from GR are observed for $\iota=0$, $q=1$ (\textit{upper left panel}), $\iota=\pi/2$, $q=1$ (\textit{upper right panel}) and $\iota=0$, $q=8$ (\textit{lower left panel}). However, a strong deviation from GR is observed the case with $\iota=\pi/2$, $q=8$ (\textit{lower right panel}) when higher modes are included in the injected signal.}
\label{fig:example_plot}
\end{figure*}

\section{Results}
\label{sec:results}

\noindent In this section, we show how the higher modes present in our injected signals, which are omitted by the GR sectors of our templates, can be interpreted as effects due to physics beyond GR. Whether these violations are observed or not will depend on the strength of the higher modes present in the injected signal, which in turn depends on the parameters of the corresponding source. Stronger higher modes will have a larger chance of being interpreted physics beyond GR, while signals for which higher modes are weak will be well recovered by the GR sectors of our templates. 

\subsection{A First Example}

\noindent The upper panels of Fig.~\ref{fig:example_plot} show the results of the parametrized test applied on signals emitted by an equal-mass binary with $M=200M_{\odot}$, for inclinations $\iota=0$ and $\iota=\pi/2$. In particular, we show the posterior distributions obtained for each of the nonGR parameters $\delta\alpha_i$. In these cases the impact of higher modes in the signal is small and the GR sector of our templates, given by the PhenomP model, can recover most of the power of the signal. This leaves almost no power available to be absorbed by the nonGR sector of the templates. Therefore, no deviations from GR are reported by the test.

In the bottom case, we set the mass ratio of the binary to $q=8$. In the face-on case, higher modes have a minimal effect and no deviations from GR are observed. However, in the edge-on case, the strong features introduced by higher modes cannot be recovered by the incomplete GR sector. Instead, an important fraction of this is mimicked by the nonGR sector of the templates. As a consequence, none of the posterior distributions includes the GR value $\delta\alpha_i =0$, and the test reports a strong deviation from GR. In all cases the simulated source has been placed at a distance such that its optimal signal-to-noise ratio (SNR) is $\rho_{\textrm{opt}}=15$ (see Eq. \ref{eq:opt_snr})

\begin{figure*}
\centering
\includegraphics[width=2\columnwidth]{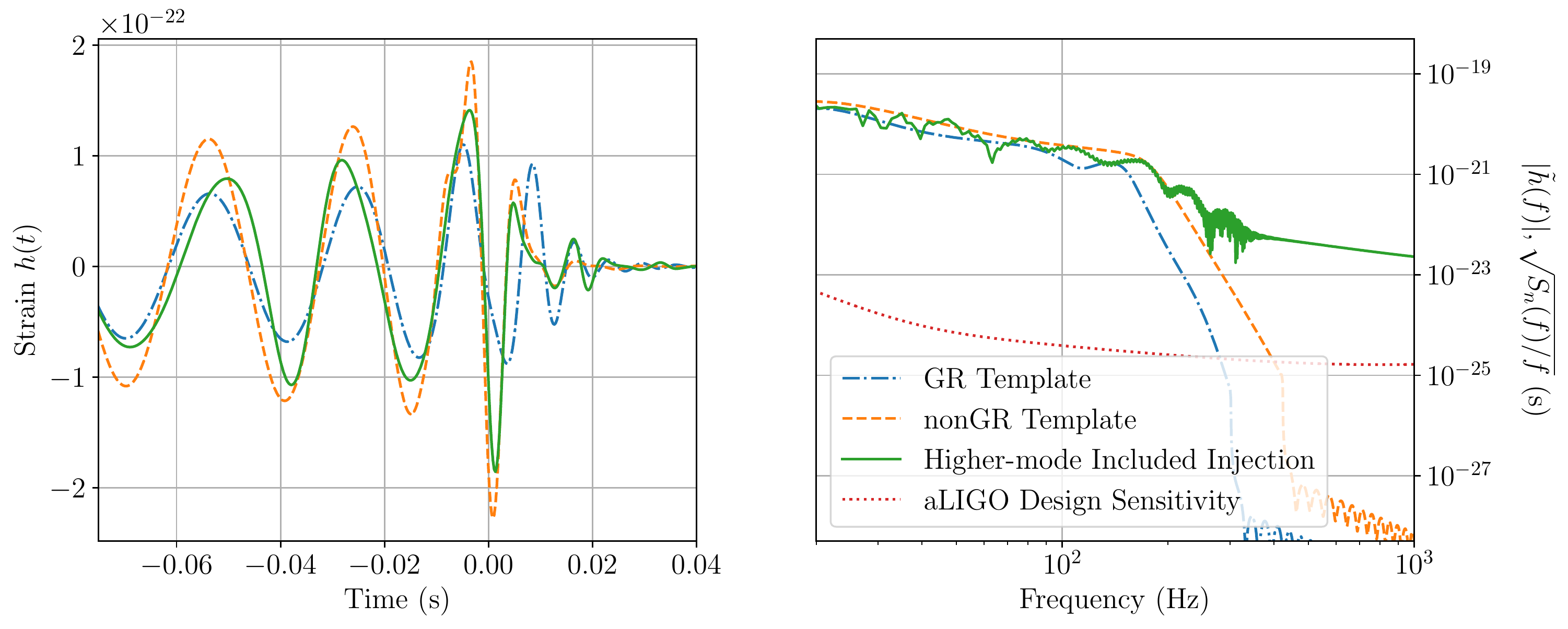}
\caption[]{\textbf{Physics beyond GR mimicking the effect of higher modes:} We show the best-fit waveform with (orange) and without (blue) testing GR parameters on top of the injected waveform (green). The latter corresponds to a BBH with $q\approx 5.5$, $\iota\approx\pi/2$. The waveforms are plotted in frequency domain (\textit{left panel}) and time domain (\textit{right panel}). Note, in the right panel, how at low frequencies both the GR and beyond GR templates show a similar performance. However, the high frequency content introduced by higher modes can only be recovered by the template enabling physics beyond GR. The right panel shows that most of these differences come from the strong effects that higher modes have in in the late merger and ringdown stages.}
\label{fig:bestfit_waveform}
\end{figure*}

Fig.~\ref{fig:bestfit_waveform} illustrates the above effect. The left panel shows, in the Fourier domain, the injected signal (green), the GR PhenomP template that best recovers it (blue) and the best fitting waveform allowing for deviations from GR (orange). The GR template is unable to capture the high frequency components introduced by the higher modes of our injection. Instead, contributions from higher modes can be accounted for when one allows for deviations from GR. The right panel shows the same waveforms in the time-domain. In this case, it is particularly clear that the differences between the performance of the GR and beyond GR templates is caused by the strong impact that higher modes have in the late merger-ringdown part of the signals. In particular, note how the blue GR waveform fails to capture the morphology of our injected signal (green) while this is quite well mimicked by when we allow for deviations from GR (orange). 

\subsection{Predicting False Violations}

\noindent As suggested in \cite{Vallisneri:2012qq}, the fractional amount of power in the injections that can not be recovered by the GR sector, and which could instead be picked up by the nonGR one, is the key ingredient to estimate the magnitude of potential deviations from GR. In our case, this can be given by the fraction of SNR that the best matching PhenomP template can recover from our injection \cite{Apostolatos:1995pj}.

The optimal SNR ratio of an injection $h_{\textrm{inj}}$ is given by 
\begin{equation}
\label{eq:opt_snr}
\rho_{\textrm{opt}}=\sqrt{(h_{\textrm{inj}} | h_{\textrm{inj}})}.
\end{equation}
Instead, the SNR that a template $h$ can recover from $h_{\textrm{inj}}$ is proportional to their overlap $\mathcal{O}(h_{\textrm{inj}}|h)$, as:
\begin{equation}
\begin{aligned}
\rho(h_{\textrm{inj}}|h)=\frac{(h_{\textrm{inj}} | h)}{\sqrt{( h | h)}} &= \frac{(h_{\textrm{inj}} | h)}{\sqrt{(h_{\textrm{inj}} | h_{\textrm{inj}})( h | h)}}\times \rho_{\textrm{opt}} \\ & \equiv \mathcal{O}(h_{\textrm{inj}}|h)\times \rho_{\textrm{opt}}
\end{aligned}
\end{equation}
The \textit{effectualness} $\cal E$ of a template family (or a template bank) to an injection $h_{\textrm{inj}} (\vec{\lambda}_{\textrm{inj}})$ is defined as the maximum overlap between $h_{\textrm{inj}} (\vec{\lambda}_{\textrm{inj}})$ and any template $h(\lambda)$ belonging to a given family, maximised over the parameters $\lambda$ \cite{Apostolatos:1995pj}. In our case:
\begin{equation}
{\cal E} = \max_{\lambda} \mathcal{O}(h_{\textrm{inj}}(\vec{\lambda}_{\textrm{inj}})|h (\vec{\lambda})) 
\end{equation}
The effectualness gives the maximum SNR (or power) from the signal that can be recovered by the non-modified PhenomP templates. The maximum SNR that can be picked by its nonGR sector is given by 
\begin{equation}
\rho_{\perp}=\rho_{\textrm{opt}} \times \sqrt{1-{\cal E}^2}.
\end{equation}

Computing $\cal E$ implies a rather expensive maximization of $\mathcal{O}$ over all the possible parameters $\vec{\lambda}_{\textrm{inj}}$. Instead, we find that a good proxy for this quantity is given by the \textit{faithfulness} of PhenomP to our injection. This is defined as the overlap of the PhenomP template having the same parameters $\vec{\lambda}_{\textrm{inj}}$ as our injection :
\begin{equation}
F=\max_{\phi_c,t_c}\mathcal{O}(h_{\textrm{inj}}(\vec{\lambda}_{\textrm{inj}}) | h (\vec{\lambda}_{\textrm{inj}})).
\end{equation}

\begin{figure*}[hbt!]
\centering
\includegraphics[width=\textwidth]{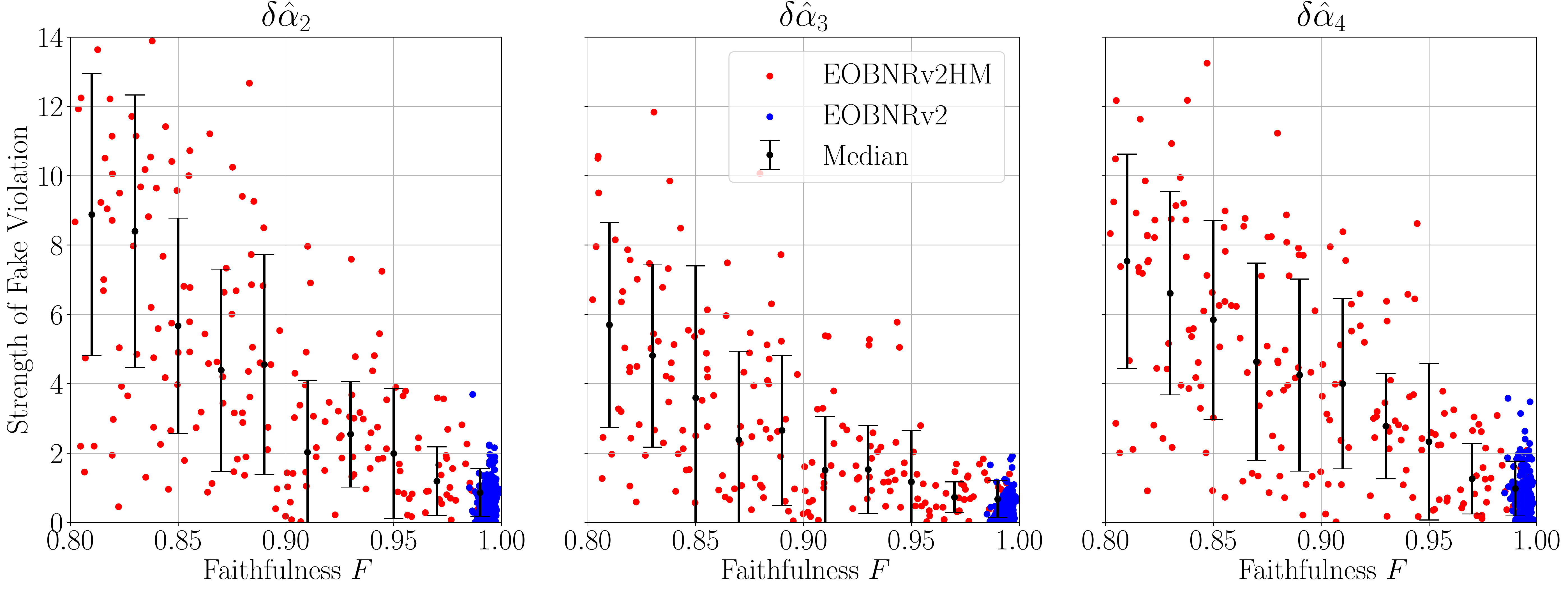}
\caption[Effect of F on TGR]{\textbf{Strength of deviations from GR as a function of faithfulness:} The panels show the strength of deviations from GR as a function of the faithfulness between our injections and the recovery PhenomP templates. The injections are distributed uniformly in co-moving volume with their optimal SNRs $\rho_{\textrm{opt}}$ distributed between 15 and 25. Note how when higher modes are included in the injections (red dots), the strength of the deviations from GR grows as the faithfulness decreases. When we omit the higher modes in the injections (blue dots) we always obtain a high faithfulness and hence, we don't observe any strong evidence for deviations from GR.}
\label{fig:dgr_match}
\end{figure*}

\begin{figure*}[hbt!]
\includegraphics[width=\textwidth]{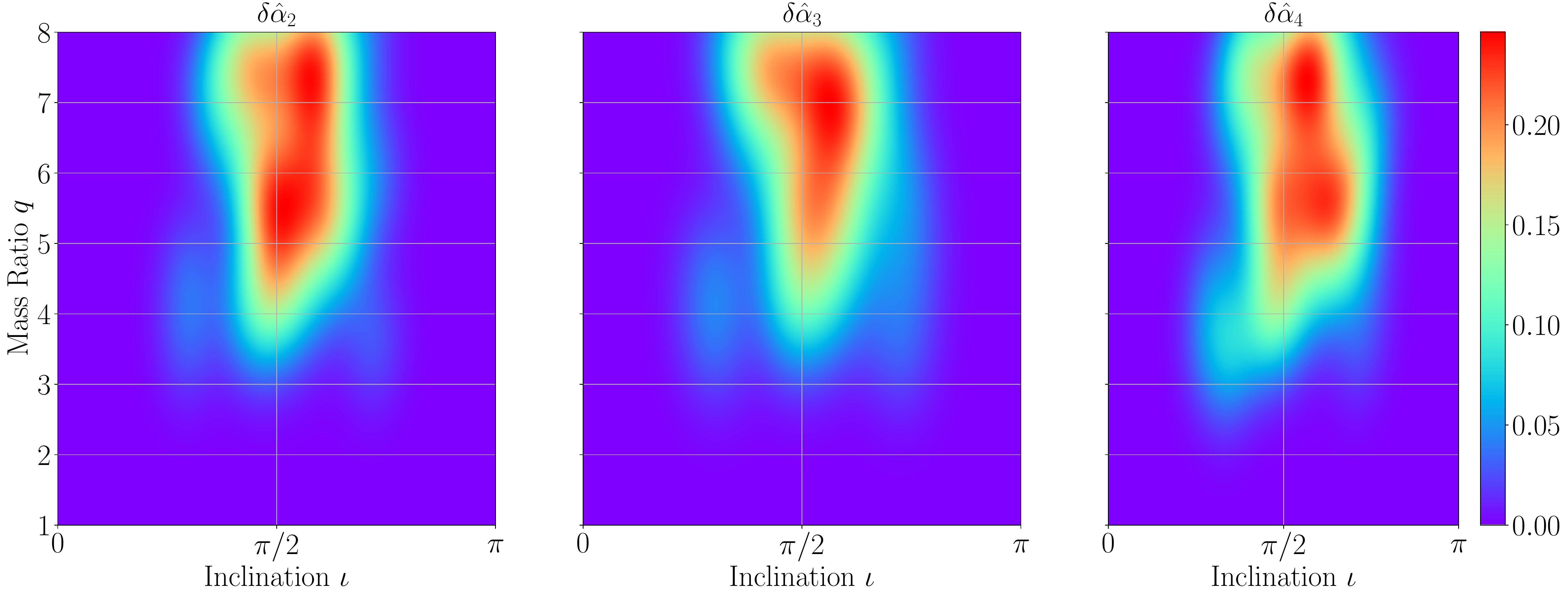}
\caption[Effect of q on TGR]{\textbf{Strength of deviations from GR as a function of source parameters: }The three panels show density of sources showing violations of GR stronger than $5\sigma$ as a function of the mass ratio $q$ and the inclination $\iota$ of the source when our injections include higher modes. Significant deviations are more likely to happen for sources with mass ratio $q > 4$ and inclination close to $\iota = \pi/2$, for which higher modes are the strongest.}
\label{fig:q_iota}
\end{figure*}

Fig.~\ref{fig:dgr_match}, shows the distribution of the strength of deviations from GR as a function of $F$. Since injections with low $\rho_{\textrm{opt}}$ will have relatively low $\rho_{\perp}$, we expect such injections to reveal no clear violations of GR. For this reason, we choose to distribute our injections uniformly in co-moving volume with $\rho_{\textrm{opt}} \in [15,25]$. \footnote{In principle, we could have extended our range down to $\rho_{opt}=0$. However, due to the high computational cost of this study, we preferred to avoid performing runs for which the result is fairly predictable.}

Red dots correspond to cases in which higher modes are included in the injections. Blue dots correspond to injections having the same parameters as the previous ones but omitting the higher modes. For the latter, we always get $F \sim 1$, and all cases can be well recovered by PhenomP, leading to no significant deviations from GR. Addition of higher modes causes $F$ to deviate from $1$ for those cases in which higher modes are strong, leaving a larger power in the injection that can be absorbed by the nonGR sector of our templates. This leads to deviations from GR whose magnitude increases, in average, as $F$ decreases. Of course, not all the power that is not recovered by PhenomP will be recovered the nonGR sector of the template. Instead, this will be proportional to the overlap between the portion of the injected waveform not recovered by the GR sector and the corresponding nonGR sector, which will vary depending on the specific effect that higher-order modes have on the injection.
As a result one may write the signal portion causing the apparent violation of GR as
\begin{equation}
h^\textrm{apparent}_\textrm{nonGR} = \rho(h_\textrm{inj} - h^\textrm{incomplete}_\textrm{GR}|h_\textrm{nonGR}) \hat{h}_\textrm{nonGR}, 
\end{equation}
with $\hat{h}_\textrm{nonGR}=h_\textrm{nonGR}/ \sqrt{(h_\textrm{nonGR}|h_\textrm{nonGR})}$.
For some of the cases we considered, the portion of signal not recovered by PhenomP is almost orthogonal to the nonGR sector of the templates, thus yielding no evidence for violations of GR. This, together with the fact that signals are injected in different realizations of Gaussian noise, is partly responsible for the spread in the strength of the violations we see for a given value of $F$. 
\begin{figure}
\centering
\includegraphics[width=\columnwidth]{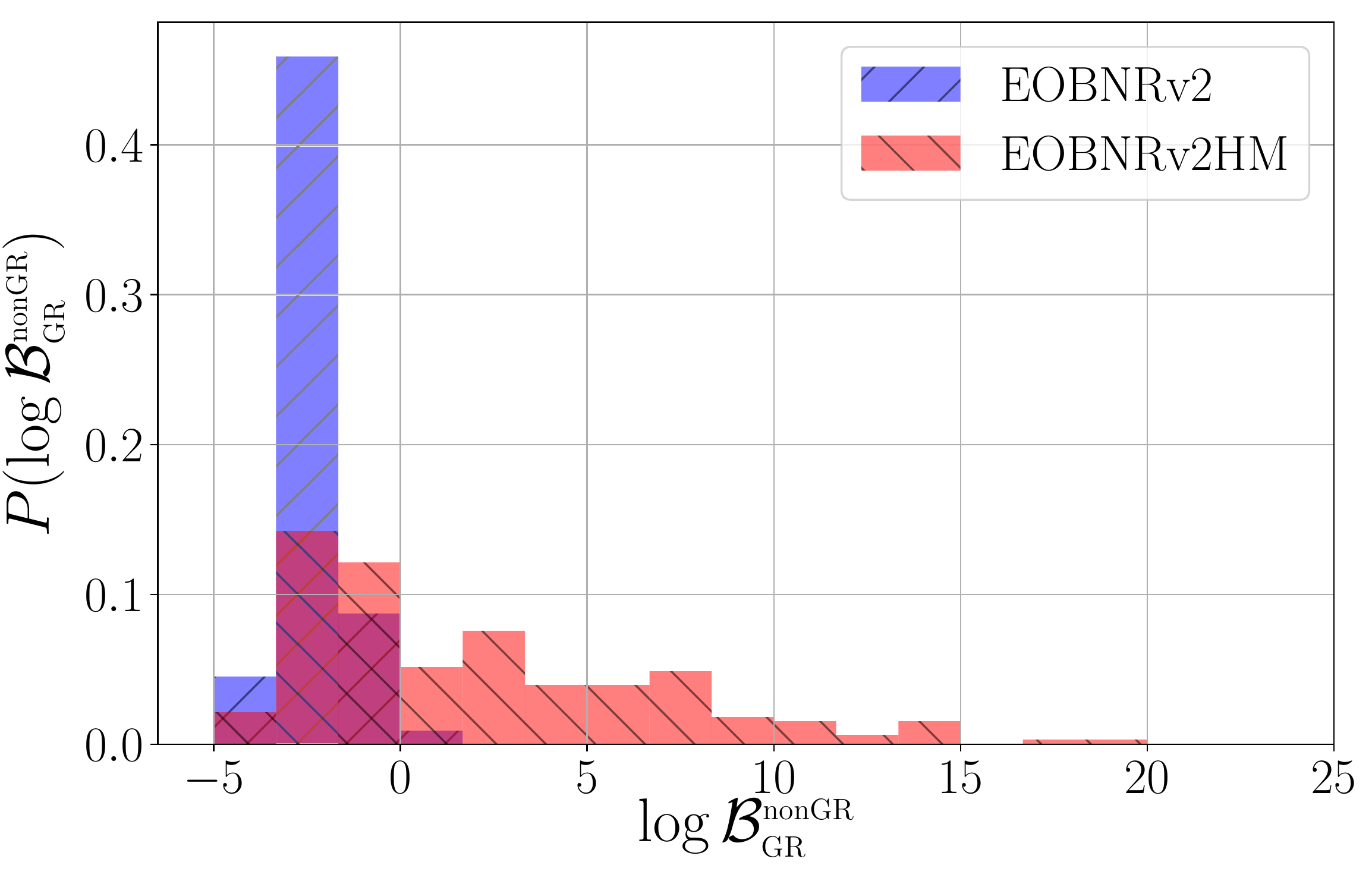} 
\caption[]{\textbf{Impact of higher modes on model selection:}  Distribution of the deviation Bayes factors for merger-ringdown parameters $\delta\alpha_i$ from 200 samples drawn from a realistic astrophysical distribution (see text for details). Injections with (EOBNRv2HM) and without (EOBNRv2) higher modes are shown in blue and red, respectively. Injections with higher modes included have the nonGR hypothesis $\mathcal{H}_{\rm nonGR}$ outrank the GR hypotheses  $\mathcal{H}_{\rm GR}$ in a significant fraction of injections.}
\label{fig:testingGR-alpha-par}
\end{figure}

\subsection{Landscape of False Violations}

\noindent Fig. \ref{fig:q_iota} shows the density of sources showing  violations of GR stronger than $5\sigma$ as a function of the inclination $\iota$ and mass-ratio $q$ of the binary. Not surprisingly, these are more likely to  occur for the cases for which higher modes are known to be stronger. This is, highly asymmetric (large $q$) binaries with an orientation far from face-on $(\iota=0)$. For these cases, the strong higher-order mode content of the signals causes the GR sector of the recovery templates (described by the PhenomP model) not being able to recover all the power from the injection, leaving a lot of power available to be picked up by the nonGR sector of the templates \cite{Pekowsky:2012sr,Varma:2016dnf,Bustillo:2016gid}. Conversely, we never observe any violation of GR for the case of face-on/off binaries or for binaries with $q < 3$.

Although the detection of a single violation from GR stronger than $5\sigma$ might be sufficient to question the correctness of GR, it is natural to ask how this evidence can accumulate with several detections, even if these are weaker than $5\sigma$. To address this, we build a set of injections with more astrophysically motivated parameters and perform parametrized tests of GR on them. These distribute uniformly in $M$, $q$ and $\cos(\iota)$. Fig. \ref{fig:testingGR-alpha-par} shows the distribution of the logarithm of the Bayes factor ${B}^{\textrm{nonGR}}_{\textrm{GR}}$ for each individual parameterized deviation for our injection set. As before, the red histogram corresponds to the case in which the injections contain higher modes while these are omitted in the blue histograms. As expected, the latter histogram shows no evidence for for deviations from GR (i.e, $\log{B} < 0$) while clear excursions toward positive values are obtained when our injections contain higher modes.

\begin{figure*}
\centering
\includegraphics[width=\textwidth]{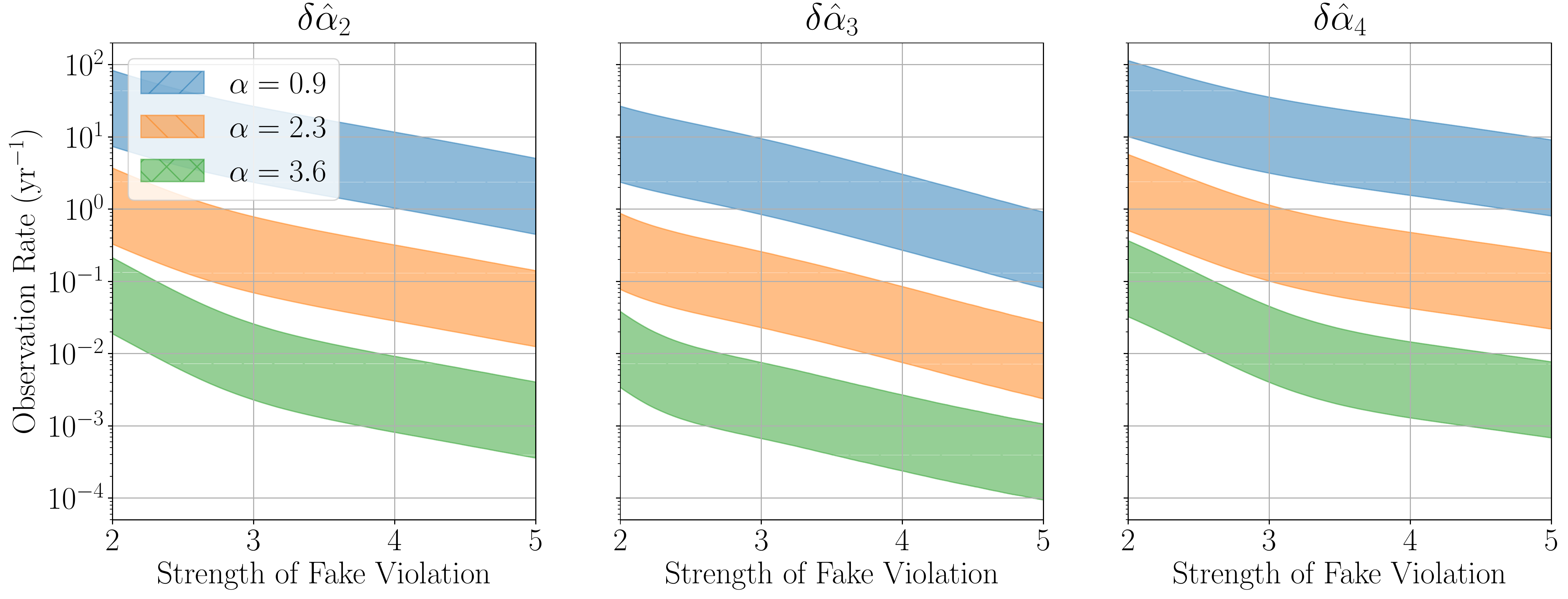}
\caption[]{\textbf{Expected observation rate of false violations of GR with a given strength:} The expected observation rate of a detection with a false violations of GR greater than a given strength for three source populations with different power law coefficients $\alpha$, where $p(m_1,m_2) \propto m_1^{-\alpha}/(m_1-5M_\odot)$, and with merger rates obtained from Ref.~\cite{PhysRevX.6.041015}. The injections are distributed uniformly in co-moving volume with their optimal SNRs distributed between $8$ and $30$. The strength is measured by the number of standard deviations $\sigma$ between the median of the posterior distribution of the testing GR parameter and the GR value ($\delta p_i = 0$). The plots report values for the testing GR parameters $\delta\hat{\alpha_2}$ (left panel), $\delta\hat{\alpha_3}$ (middle panel) and $\delta\hat{\alpha_4}$ (right panel). For the flattest power law ($\alpha=0.9$), an observation of a $5\sigma$ deviation from GR is expected to occur around once per year.}
\label{fig:dGRNT}
\end{figure*}

\subsection{Rate of False Violations of GR}

\noindent To close this section, we address the question of how long would it take to observe a fake violation of GR using current parametrized tests. To assess this we consider a source distribution more realistic than the one in the previous paragraph. This is driven by the recent population distributions and merger rates of BBH recently obtained by the LSC. In particular, we consider the same standard BBH population used by the LSC in \cite{PhysRevLett.118.221101}: a power law distribution $m_1$ and uniform in $m_2$ $p(m_1,m_2) \propto m_1^{-\alpha}/(m_1-5M_\odot)$ and consider the merging rates derived from the LIGO First Observation Run \cite{PhysRevX.6.041015} and GW170104\cite{Abbott:2017vtc}. We distribute our injections uniformly in co-moving volume with their optimal SNRs $\rho_{\textrm{opt}}$ ranging between $8$ and $30$. With these ingredients, we can compute the time that it would take to ``detect'' a signal $h_{\textrm{inj}}$ such that performing a parametrized test on it would lead to a false violation stronger than $N \sigma$. For our purposes, ``detect'' means that the corresponding source would be observed with an SNR $\rho_{\textrm{obs}} \geq 8$, which is the standard threshold used in the literature. Moreover, we assume that the search observing the event can recover its full optimal SNR $\rho_{opt}= \sqrt{(h_{\textrm{inj}}|h_{\textrm{inj}})}$. Hence, we only need to compute how many sources show strong fake deviations and coalesce close enough to us to produce $\rho_{\textrm{opt}}\geq 8$ per unit time. 

We note that assuming $\rho_{\textrm{obs}} = \rho_{\textrm{opt}}$ may be too optimistic. In fact, matched filter searches are designed to recover $\rho_{\textrm{obs}} \geq 0.956 \rho_{\textrm{opt}}$ in the regions of the parameter space that are well covered by their template banks and, more importantly, these are not currently sensitive enough to signals with strong higher modes \cite{Usman:2015kfa,Nitz:2017svb}. However, promising progress has been made to make these searches sensitive to such signals \cite{Harry:2017weg}. Furthermore, more generic searches which remain agnostic to the morphology of the signals \cite{Klimenko:2008fu} should ideally be able to recover all of the optimal SNR of any kind of signal. 

Fig. \ref{fig:dGRNT} shows the expected observation time for observing a false violation of GR with given strength coming from each of the considered nonGR parameters, and assuming a three detector network formed by advanced LIGO and advanced Virgo operating at their design sensitivities. Results are shown for source populations corresponding to three different values of $\alpha$. In particular, these are chosen to be the 5th, 50th and 95th percentile of the $\alpha$ distribution inferred in \cite{PhysRevLett.118.221101}. The width of the bands corresponds to the 90$\%$ credible interval of the event based merger rate.

We note again that for a signal to show strong false violations of GR, it requires a strong higher order mode content. Hence, the corresponding source should have a large mass ratio and be fairly edge-on. Signals from these sources are much weaker than signals from nearly equal-mass ones, or face-on sources. It is therefore reasonable that the vast majority of current and future BBH detections will mostly correspond to the latter group, reducing the chances of observing false violations from GR produced by higher modes.

Consistently, in the worst case, corresponding to the source distribution having $\alpha=0.9$, we estimate that a fake violation with a significance of $5\sigma$ can occur once per year (left and right panels) of observation. Although the probability of observing this kind of effect seems quite low, a single of these observations might be enough to put into question the correctness of the theory of General Relativity.

\section{Conclusions}
\label{sec:conclusions}

\noindent The direct detection of gravitational waves from the coalescence of binary black holes has allowed for tests General Relativity in the strong gravitational field regime. So far, the results of all tests are consistent with General Relativity being correct. In this paper, we have focused on one of these tests, known as parameterized test, which seeks evidence for physics beyond General Relativity in the incoming signals. To this, gravitational wave signals are compared to waveform templates that account for possible deviations from General Relativity. In doing this, it is crucial that part ( or ``sector'') of these waveforms that accounts for physics within General Relativity includes all of its effects. Otherwise, missing physics might be mimicked by the non-GR sector of these templates. This would lead the analysis to report false apparent violations of General Relativity.

Current parameterized tests implement templates which omit two physical phenomena present in General Relativity: the eccentricity of the orbit of the binary, and the so called higher-order modes of the gravitational wave emission. This is partly due to the lack of waveform models accounting for both higher-order modes and generic black hole spins. However, we note that waveform models including higher modes and aligned spin effects are under active development \cite{London:2017bcn}.

In this work, we have demonstrated that when higher-order modes are present in the signal but are omitted in the templates, its effect can be mimicked by physics beyond General Relativity. This leads to apparent violations. To this end, we have performed parametrized tests on simulated signals containing higher-order modes. We have done this for a large family of non-spinning binaries with total masses ranging in $[100,200]M_\odot$, mass ratios $q \leq 8$ and varying inclinations. Strong deviations from General Relativity, with significance larger than $5\sigma$, are reported for highly asymmetric binaries $(q \geq 4)$ whose orbital angular momentum is almost orthogonal to the line-of-sight (or edge-on oriented). This is consistent these sources being known to have a strong higher-order mode emission.

Finally, considering several astrophysically motivated source distributions and assuming an Advanced LIGO - Advanced Virgo network operating at design sensitivity, we have computed the rate at which false violations could arise if current parametrized test of GR are performed in future detections. In the most worst scenario, we estimate that false observations from GR could be observed once per year.

In this study we have focused on parametrized tests of General Relativity. However, there is a whole family of similar tests which involve the comparison of data with templates that are modified to account for given effects beyond General Relativity. Examples of these are searches for dipole radiation or tests of the mass of the graviton. It is reasonable to expect that the effects described in this work do affect these and similar studies. Hence, we recommend that in the future,  effects beyond General Relativity is are added to templates that account for higher-order modes, and preferably also eccentricity.
Finally, we stress that the low mass ratio of the binary black holes detected by Advanced LIGO and Virgo so far, makes it unlikely that the effects we have described could affect the tests of GR based on these observations.

\section{Acknowledgements}
\noindent The authors are grateful for useful comments by Salvatore Vitale, Parameswaran Ajith, Chris Van Den Broeck and Abbay Ashtekar.
JCB gratefully acknowledges support from the NSF grants 1505824, 1505524, 1333360. 
PTHP, YFW and TGFL were partially supported by a grant from the Research Grants Council of Hong Kong (Project No. CUHK 24304317), and the Direct Grant for Research from the Research Committee of the Chinese University of Hong Kong.

\bibliography{IMBBH}
\end{document}